\documentclass[sigconf,authorversion,nonacm]{acmart} 
\acmBooktitle{Proceedings of the 32nd ACM Symposium on the Foundations of Software Engineering (FSE '24), November 15--19, 2024, Porto de Galinhas, Brazil}

\usepackage{amsmath}
\usepackage{algorithmic}
\usepackage{graphicx}
\usepackage{textcomp}
\usepackage{xspace}
\usepackage{hyperref}
\usepackage[capitalise,nameinlink,noabbrev]{cleveref}
\usepackage{alltt}
\usepackage[caption=false]{subfig}
\usepackage{boxedminipage}
\usepackage{multicol}
\usepackage{booktabs}
\usepackage{listings}

\usepackage{tikz}
\usetikzlibrary{tikzmark}
\usetikzlibrary{arrows.meta}

\usepackage{relsize}  % \smaller
\def\ALHAZEN{{\smaller{}ALHAZEN}\xspace}
\def\AVICENNA{{\smaller{}AVICENNA}\xspace}
\def\SFLKIT{{SFLKit}\xspace}

\def\FUZZBENCH{{FuzzBench}\xspace}
\def\CODEFLAWS{{Codeflaws}\xspace}
\def\DEFECTS4J{{Defects4J}\xspace}
\def\BUGSJS{{BugsJS}\xspace}
\def\BUGSINPY{{BugsInPy}\xspace}
\def\TESTS4PY{{Tests4Py}\xspace}

\def\PYTEST{{\smaller{}PYTEST}\xspace}
\def\REFACTORY{{Refactory}\xspace}
\def\BUGSWARM{{BugSwarm}\xspace}
\def\BEARS{{Bears}\xspace}
\def\BUGSJAR{{Bugs.jar}\xspace}
\def\COOKIECUTTER{{Cookiecutter}\xspace}
\def\FASTAPI{{FastAPI}\xspace}
\def\HTTPIE{{HTTPie}\xspace}
\def\SANIC{{Sanic}\xspace}
\def\PYSNOOPER{{PySnooper}\xspace}
\def\THEFUCK{{The Fuck}\xspace}
\def\YOUTUBEDL{{youtube-dl}\xspace}
\def\MIDDLE{{middle}\xspace}
\def\MARKUP{{markup}\xspace}
\def\CALCULATOR{{calculator}\xspace}
\def\EXPRESSION{{expression}\xspace}

\definecolor{rltred}{rgb}{0.5,0,0}
\definecolor{rltgreen}{rgb}{0,0.5,0}
\definecolor{rltblue}{rgb}{0,0,0.5}
\hypersetup{
colorlinks=true,
urlcolor=rltblue,
linkcolor=rltred,
citecolor=rltgreen,
}

\usepackage[colorinlistoftodos,textsize=tiny]{todonotes}
\newcommand{\maketodo}[2]{\expandafter\newcommand\csname #1\endcsname[1]{\todo[bordercolor=#2!80!black,color=#2]{\textbf{\MakeUppercase#1:} ##1}\xspace}}

\presetkeys{todonotes}{fancyline}{}
\definecolor{todored}{rgb}{1, 0.6, 0.6}
\definecolor{todoorange}{rgb}{1, 0.8, 0.4}
\definecolor{todoblue}{rgb}{0.4, 0.8, 1.0}
\definecolor{todogreen}{rgb}{0.8, 1.0, 0.4}
\definecolor{todopurple}{RGB}{255, 100, 127}

\maketodo{NOTE}{todogreen}
\maketodo{TODO}{todored}
\maketodo{CHECK}{todoblue}
\newcommand{\DONE}[1]{} % Older arguments, now addressed
\newcommand{\WONTFIX}[1]{} % Other arguments, won't be addressed
\newcommand{\ALSO}[1]{} % Additional arguments

\usepackage{syntax}  % grammars
\def\nonterm#1{\textnormal{$\langle$\emph{#1}$\rangle$}}

\newcommand{\mathid}[1]{\textit{#1}}
\newcommand{\codeid}[1]{\texttt{#1}}
\let\backslashpipe=\|
\let\backslashstar=\*  % in case we ever need \*
\let\backslashangle=\<
\def\|#1|{\mathid{#1}}  % Math Identifier
\def\*#1*{\codeid{#1}}  % Code
\def\<#1>{\nonterm{#1}} % Nonterminal

\usepackage{colortbl}

\definecolor{DarkBlue}{rgb}{0.0859, 0.308, 0.523}
\definecolor{DarkOrange}{rgb}{0.8, 0.4, 0.0}
\definecolor{DarkGreen}{rgb}{0.00,0.40,0.00}
\definecolor{ScarletRed}{rgb}{0.60,0.00,0.00}
\definecolor{AlmostWhite}{rgb}{0.80,0.80,0.80}
\definecolor{Gray}{gray}{0.85}
\definecolor{Cornsilk}{rgb}{0.98, 0.94, 0.9}

{\medskip
\noindent
\let\emph=\textbf
\begin{boxedminipage}{\linewidth}\begin{center}\em}%
{\end{center}\end{boxedminipage}%
\medskip
}

\newsavebox{\mybox}
\usepackage{tcolorbox}
\tcbuselibrary{listings,skins}

\lstdefinestyle{mystyle}{
  xleftmargin=0pt,
   basicstyle={\footnotesize\ttfamily},
   aboveskip=3mm,
   belowskip=3mm,
   keywordstyle=\bfseries,
   showstringspaces=false,
  escapechar=?,
  language=Python
}
\definecolor{code_indent}{HTML}{CCCCCC}

\newcommand{\indentrule}{\color{code_indent}\vrule\hspace{2pt}}

\newtcblisting{mylisting}[2][]{
  arc=0pt, outer arc=0pt,
  listing only,
  listing style=mystyle,
  title={\normalsize #2},
  #1
}

\author{Marius Smytzek}
\orcid{0000-0002-4899-9031}
\affiliation{%
	\institution{CISPA Helmholtz Center for Information Security}
  	\city{Saarbr{\"u}cken}
	\country{Germany}
}
\email{marius.smytzek@cispa.de}

\author{Martin Eberlein}
\orcid{0000-0003-4268-7632}
\affiliation{%
	\institution{Humboldt-Universit{\"a}t zu Berlin}
  	\city{Berlin}
	\country{Germany}
}
\email{martin.eberlein@hu-berlin.de}

\author{Batuhan Ser\c{c}e}
\orcid{0009-0007-5353-6318}
\affiliation{%
	\institution{CISPA Helmholtz Center for Information Security}
  	\city{Saarbr{\"u}cken}
	\country{Germany}
}
\email{batuhan.serce@cispa.de}

\author{Lars Grunske}
\orcid{0000-0002-8747-3745}
\affiliation{%
	\institution{Humboldt-Universit{\"a}t zu Berlin}
 	\city{Berlin}
	\country{Germany}
}
\email{grunske@hu-berlin.de}

\author{Andreas Zeller}
\orcid{0000-0003-4719-8803}
\affiliation{%
	\institution{CISPA Helmholtz Center for Information Security}
  \city{Saarbr{\"u}cken}
	\country{Germany}
}
\email{zeller@cispa.de}

\title{Tests4Py: A Benchmark for System Testing}

\begin{abstract}
  Benchmarks are among the main drivers of progress in software engineering research. However, many current benchmarks are limited by inadequate system oracles and sparse unit tests.
  Our \emph{\TESTS4PY{}} benchmark, derived from the \emph{\BUGSINPY{}} benchmark, addresses these limitations.
  It includes 73 bugs from seven real-world Python applications and six bugs from example programs.
  Each subject in \TESTS4PY{} is equipped with an oracle for verifying functional correctness and supports both system and unit test generation.
  This allows for comprehensive qualitative studies and extensive evaluations, making \TESTS4PY{} a cutting-edge benchmark for research in test generation, debugging, and automatic program repair.

\end{abstract}



\begin{document}

\maketitle

\begin{figure}
	\includegraphics[width=\columnwidth]{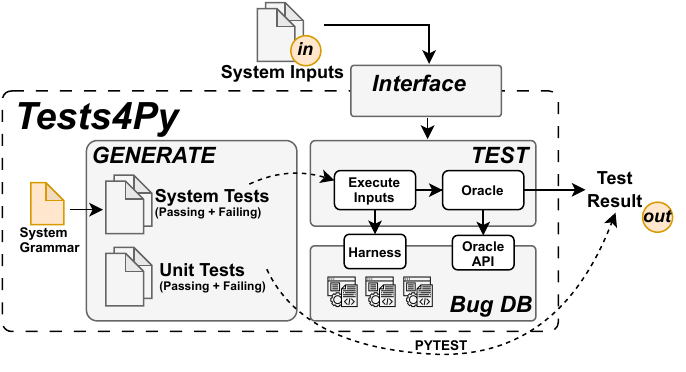}
	\vspace{-2\baselineskip}
	\caption{\TESTS4PY{} Overview. \TESTS4PY incorporates components for generating system and unit tests, running them, and assessing their results using generic oracles.}%
	\label{fig-overview}
	\vspace{-\baselineskip}
\end{figure}

\section{Introduction}%
\label{sec-introduction}

For several years, benchmarks of program faults have been the backbone for evaluating methodologies and driving qualitative studies in the domain of software engineering research \cite{widyasari2020bugsinpy, metzman2021fuzzbench, rene2014defects4j}.
These benchmarks aim to allow engineers to investigate faults within a real-world application context.
Consequently, these benchmarks predominantly consist of programs with identifiable faults, paving the way for thorough investigations of bugs.

Google's \FUZZBENCH{}~\cite{metzman2021fuzzbench}, for example, is a service to evaluate fuzz testing tools on various real-world subjects.
Similarly, \CODEFLAWS{}~\cite{tan2017codeflaws} is a benchmark developed for automatic program repair, wherein bugs are sorted into categories to yield insights into the types of bugs that can be repaired.
Beyond benchmarks with targeted goals, numerous benchmarks are tailored for specific programming languages.
Notable examples include \DEFECTS4J{}~\cite{rene2014defects4j}, \BUGSJS{}~\cite{gyimesi2019bugsjs}, and \BUGSINPY{}~\cite{widyasari2020bugsinpy} for Java, JavaScript, and Python, respectively.

Despite these benchmarks' pivotal role in software engineering research, the complexity of automated debugging and test generation approaches often requires more comprehensive evaluations.
For instance, test generation benchmarks currently rely on \emph{generic} oracles, such as crash detection, which, while being practical for gauging program security, could benefit from evaluations that determine their efficacy in uncovering \emph{functional} bugs.
Detecting functional bugs requires test generators capable of producing \emph{oracles}, which remains a research challenge.

Another significant limitation is that benchmarks typically offer a set of unit tests with \emph{fixed inputs,}
lacking interfaces to incorporate \emph{generated inputs.} This restricts the potential for exploring combinations of test generators and automated repair tools.

Our novel benchmark, \TESTS4PY{}, addresses and overcomes these limitations for Python programming.
\TESTS4PY{} (\Cref{fig-overview}) is an extendable suite that consists of various faults, each meticulously sourced from five real-world Python programs.
These bugs are adopted from the \BUGSINPY{} database and are thoughtfully augmented with an \emph{oracle,} a \emph{system interface,} and the capability to \emph{incorporate both system and unit test generators.}

A key characteristic of \TESTS4PY{} is its emphasis on \emph{test diversity} to foster a more extensive and rigorous evaluation.
Therefore, each faulty program included in our benchmark is coupled with a comprehensive set of carefully constructed system and unit tests that can be used to guarantee test diversity if needed.
Half of these tests are designed to pass successfully, while the other half are crafted to fail, simulating various scenarios.
This unique design balances capturing potential faults and affirming the program's functionality, providing a comprehensive and practical benchmarking tool in the Python programming environment.

To illustrate \TESTS4PY{}, let us consider bug~\#2 from \FASTAPI{}~\cite{fastapi},\footnote{\TESTS4PY{} uses the same bug identifiers as \BUGSINPY{}.} which occurs when \FASTAPI{} establishes a WebSocket while overriding its dependencies.
Each bug in our benchmark comes with at least one failing unit test. The unit test for \FASTAPI{} bug~\#2 (\Cref{fig-unittest}) is included in the \FASTAPI{} project and was adopted by \BUGSINPY{}.
This particular unit test includes an oracle specific to this test case.
Thus, it cannot be used in conjunction with \emph{generated tests} (as it has only one input and related oracle), and it cannot be used with \emph{system inputs} (given or generated).
This is where \TESTS4PY{} steps in:
\begin{itemize}
    \item First, \TESTS4PY{} offers an \emph{interface for system tests} (\Cref{fig-api}) that runs the project through a custom \emph{harness} that is included in the benchmark.
    \item Second, \TESTS4PY{} provides an \emph{oracle} that is suitable for generated inputs (\Cref{fig-oracle}).
    The {generic} oracle for \FASTAPI{} bug~\#2
    examines the run and checks for signals that indicate whether the defect has been triggered.
    \item On top, \TESTS4PY{} provides hand-crafted \emph{grammars}, specifying the input of the program to \emph{generate} and \emph{validate} further inputs (with outcomes checked through its oracles).
\end{itemize}

Assembling these oracles, tests, and test generation interfaces for each subject and bug required substantial effort and attention to detail.
We began with a deep dive into each fault, aided by unit tests and fixes from the \BUGSINPY{} database. Based on our understanding, we designed twenty unique unit tests per fault.
If a subject lacked an interface capable of triggering the fault, we implemented a harness to fill this gap, furthering the authenticity of our testing environment.
While time-intensive, this exhaustive process was integral to creating a more comprehensive benchmark.
By addressing the previously identified limitations, we aimed to enhance the value of our benchmark, \TESTS4PY{}, for the software engineering community.

With \TESTS4PY{}, we make the following contributions:%
\begin{description}%[style=sameline,topsep=0pt]  % Don't mess with standard settings -- AZ
    \item[An easy-to-use benchmark.]
    We provide a simple \emph{command line interface}. 
    We also incorporate the entire \BUGSINPY{} database, which is easily extendable with new bugs in \TESTS4PY{}.
    \item[Oracles.]
    We provide an \emph{oracle} for every subject, facilitating the verification of (given and generated) system tests, including functional testing.
    \item[Interfaces for test generation.]
    We include \emph{system and unit test generators} for each subject, enabling the creation of large and diverse test sets.
    \item[Input specifications.]
    We provide \emph{input grammars} to specify the format of the system tests for each included bug.
\end{description}

\TESTS4PY{} is available as open source; see \Cref{sec-conclusion} for details.

\begin{figure}[t!]
\begin{scriptsize}
\begin{mylisting}[left=2mm,width=\linewidth,top=0pt,bottom=0pt]{\textit{\FASTAPI{}} bug~\#2 failing unit test}
def ?{\textbf{test_router_ws_depends_with_override}}?():
?\indentrule? client = TestClient(app)
?\indentrule? app.dependency_overrides[ws_dependency]
?\indentrule? ?\indentrule? = lambda: "Override"
?\indentrule? ?\textbf{with}? client.websocket_connect("/router-ws-depends/")
?\indentrule? ?\indentrule? ?\textbf{as}? websocket:
?\indentrule? ?\indentrule? ?\indentrule? ?\textbf{assert}? websocket.receive_text() == "Override"
\end{mylisting}
\end{scriptsize}
\vspace{-0.5\baselineskip}
\caption{The original unit test for the \FASTAPI{} bug~\#2 as included in the \FASTAPI{} project.}%
\label{fig-unittest}
\end{figure}

\begin{figure}[t!]
\begin{scriptsize}
\begin{mylisting}[left=2mm,width=\linewidth,top=0pt,bottom=0pt]{\TESTS4PY{}: \textit{\FASTAPI{}} bug~\#2 interface}
def ?{\textbf{run}}?(system_test: ?\textbf{List}?[str]):
?\indentrule? subprocess.run(
?\indentrule? ?\indentrule? ["python", HARNESS_FILE] + system_test,
?\indentrule? ?\indentrule? env=execution_environment, stdout=subprocess.PIPE
?\indentrule? ).stdout
\end{mylisting}
\end{scriptsize}
\vspace{-0.5\baselineskip}
\caption{The \TESTS4PY{} interface (simplified) provides a \emph{harness and API} to execute system tests for the \FASTAPI{} bug~\#2. The result of this function gets directly provided to the oracle.}%
\label{fig-api}
\end{figure}

\begin{figure}[t!]
\begin{scriptsize}
\begin{mylisting}[left=2mm,width=\linewidth,top=0pt,bottom=0pt]{\TESTS4PY{}: Oracle for \textit{\FASTAPI{}} bug~\#2}
def ?{\textbf{oracle}}?(output) -> TestResult:
?\indentrule? if (mode == "websocket" and url in websockets 
?\indentrule? ?\indentrule? ?\indentrule? and websockets[url] in overrides 
?\indentrule? ?\indentrule? ?\indentrule? and overrides[websockets[url]] not in output):
?\indentrule? ?\indentrule? return TestResult.FAILING
?\indentrule? else:
?\indentrule? ?\indentrule? return TestResult.PASSING
\end{mylisting}
\end{scriptsize}
\vspace{-0.5\baselineskip}
\caption{The \TESTS4PY{} oracle (excerpt and abstracted) for \FASTAPI{} bug~\#2, used to validate system tests, checks for \emph{generic} issues. The input itself describes the \lstinline|mode|, \lstinline|websockets|, and \lstinline|overrides|.}%
\label{fig-oracle}
\end{figure}

%The remainder of this paper is structured as follows.
%\Cref{sec-benchmark} describes the general design of the benchmark. 
%\Cref{sec-framework} provides information about the usage of \TESTS4PY{}. 
%\Cref{sec-use-cases} provides information about various use cases for the benchmark and demonstrates its power and benefits. 
%Finally, we discuss possible threats, related work, and future work in \Cref{sec-ttv,sec-related,sec-conclusion}, respectively.

\section{TESTS4PY and its Benchmark}%
\label{sec-benchmark}

\TESTS4PY{} builds on the bugs in the existing \BUGSINPY{}~\cite{widyasari2020bugsinpy} benchmark.
This decision significantly expedited our process, alleviating the need to identify bugs from the ground up.
However, every subject we selected underwent rigorous verification against the initially provided unit tests.
Any subject that failed to reproduce the bug using unit tests was duly discarded, a fate that befell several \BUGSINPY{} subjects.
Given the extensive collection of real-world bugs already at our disposal, the unique feature of the \TESTS4PY{} benchmark is its flexibility in testing the included subjects.
Each subject must be capable of \emph{accepting inputs to enable system testing.}
% However, most of our selected subjects needed a direct interface.
For this purpose, we implemented \emph{harnesses} for virtually all subjects where direct system input was not feasible, maintaining close fidelity to the original defect.
So far, our \TESTS4PY{} benchmark includes 73~bugs sourced from seven different real-world projects: \textit{\COOKIECUTTER{}}~\cite{cookiecutter} with 3 bugs, \textit{\FASTAPI}~\cite{fastapi} with 16 bugs, \textit{\HTTPIE{}}~\cite{httpie} with 5 bugs, \textit{\PYSNOOPER{}}~\cite{rachum2019pysnooper} with 2 bugs, \textit{\SANIC{}}~\cite{sanic} with 5 bugs, \textit{\THEFUCK{}}~\cite{thefuck} with 32 bugs, and \textit{\YOUTUBEDL{}}~\cite{youtubedl} with 10 bugs.
Moreover, we implemented four example programs with six bugs to enable users to quickly and initially evaluate their setup before evaluating the real-world subject. These programs are labeled as \textit{\CALCULATOR{}}, \textit{\EXPRESSION{}}, \textit{\MARKUP{}}, and \textit{\MIDDLE{}}.

\subsection{\TESTS4PY{} Components}

\Cref{fig-overview} depicts the three-tier architecture of \TESTS4PY{}, similar to its predecessor, \BUGSINPY{}.
This architecture includes a bugs database for metadata storage, a database abstraction layer to make this information accessible, and an execution framework for testing bugs using existing or newly generated tests.
Additionally, \TESTS4PY{} facilitates test case \emph{generation}, where it can produce numerous tests with a defined ratio of failing ones.
Each test adheres to the specified input grammar. 
An oracle evaluates these tests and categorizes their results.
While adding new bugs to \TESTS4PY{} is straightforward, fully leveraging its features requires manual effort, influenced by factors like the test harness or the complexity of inputs needed.

% \Cref{fig-overview} shows the components of \TESTS4PY{}.
% Like its predecessor, \BUGSINPY{},
% \TESTS4PY{} adheres to a \emph{three-tier architecture.}
% This structure consists of (1)~a \emph{bugs database}, serving as a repository that houses the metadata for each subject; (2)~a \emph{database abstraction layer}, rendering the information from the bugs database into formats accessible by both humans and machines;
% and (3)~an \emph{execution framework} to test each bug using either the provided or generated tests, bringing the entire process full circle and enabling a comprehensive evaluation of each fault.

% The \TESTS4PY \emph{components} include the \emph{generation} of new test cases and the subsequent \emph{testing} of these cases.
% The generation process can produce an arbitrary number of tests, with a specified proportion of failing tests.
% Every generated system test aligns with the provided input specification, i.e., the grammar.
% %
% The oracle subsequently assesses these system tests.
% The oracle gets information about the executed subject, scrutinizes the observation, and classifies it as passing, failing, or unknown in cases where the oracle cannot precisely determine it.
% This meticulous process ensures a comprehensive and reliable testing framework within \TESTS4PY{}.

% While extending \TESTS4PY{} by a new bug is straightforward, fully integrating all of its capabilities requires manual effort.
% This effort depends, among others, on the design of the harness or the complexity of the required inputs.

\subsection{Oracles}%
\label{sec-oracles}

The harnesses we have implemented are pivotal in conveying the necessary information to discern a failure.
In alignment with this objective, we have incorporated a unique testing \emph{oracle} for each bug to determine if a system input incites the defect.
The oracles are diverse, reflecting the varied nature of the projects and bugs. Each subject necessitates an individual oracle, resulting in many implementations.
These oracles range from capturing the standard error stream and detecting specific exceptions to asserting the existence of files or directories and verifying the return values of particular functions.
The choice of implementation is dictated by what best aids in identifying the inherent defect.
Accessing these oracles is a straightforward process.
One can create a file with the test input and then incorporate \TESTS4PY{}'s system test module.
This design facilitates easy utilization and encourages the active use of the benchmark in varied testing scenarios.

\subsection{Grammars}%
\label{sec-specification}

When introducing system inputs, an essential element is the requirement to adhere to a specific format to ensure they match the interface of the included bugs and are not dismissed during the input parsing and processing phase.
Therefore, each subject in \TESTS4PY{} includes \emph{grammars} that serve as input specifications for the system tests.
Each grammar is valuable for validating the system tests created syntactically.
Furthermore, grammars are used within the oracles to gather additional information, helping distinguish between successful and failing test runs.
\TESTS4PY{} thus guarantees the integrity of the system inputs, increasing benchmarking validity.

\subsection{System Tests}%
\label{sec-systemtests}

By utilizing the harnesses, oracles, and grammars, \TESTS4PY{} can accurately distinguish passing and failing tests at the system level.
Furthermore, these components lay the groundwork for generating new test cases.
Not only can we validate these tests, but our thorough fault analysis also enabled the implementation of a targeted test generation for each subject.
As a result, we obtained a precise understanding of how to either trigger or avoid the defect, enhancing the precision and utility of the generated tests.
Test generation depends entirely on the specific project and the bug it targets.

The test generation is mostly hand-crafted to suit these dependencies while relying on known testing techniques like fuzzing to generate specific input components.
For the example of bug~\#2 from the \FASTAPI{} project from above, this generated input describes what different applications, routers, requests, and responses exist for the test server set up by the harness.

\subsection{Unit Tests}%
\label{sec-unittests}

While system tests are crucial for facilitating test generation, there are requirements for specific methodologies that they cannot fulfill.
For example, many automated program repair strategies heavily rely on unit tests.
To accommodate this and to ensure that each subject can be thoroughly tested, we incorporated the ability to generate and execute unit tests into our benchmark.
\Cref{fig-overview} also delineates the unit test module of our benchmark.
Instead of generating system inputs defined by a grammar, we directly produce Python code as a \texttt{unittest.TestCase}.
This generated test case is then executed by the \PYTEST{} testing framework as part of the entire test set or as individual tests.
This enhancement broadens the scope of testing, making \TESTS4PY{} more versatile and comprehensive.

\subsection{Usage}%
\label{sec-framework}

To install \TESTS4PY{}, run \verb|pip install tests4py| from the command line.
After installation, use commands like
\begin{enumerate}
	\item \verb|t4p info| to retrieve information of the included projects;
	\item \verb|t4p checkout -p FastAPI -i 2| to download bug~\#2 from the \verb|FastAPI| project;
	\item \verb|t4p build| from the generated directory to build the virtual environment and install the subject in it;
	\item \verb|t4p systemtest generate -n 10| to generate ten system tests in the newly created folder; and
	\item \verb|t4p systemtest test| to run these tests.
\end{enumerate}

\section{TESTS4PY Use Cases}%
\label{sec-use-cases}

We want to highlight and discuss several use cases for \TESTS4PY{}.

\subsection{Evaluating Test Generation}%
\label{sec-tg}
Every subject in \TESTS4PY{} comes equipped with an oracle and input specification, providing a fertile ground for evaluating test generation techniques such as grammar-based fuzzing~\cite{soremekun2022ifh, eberlein2020evogfuzz}.
Unlike previous benchmarks that relied on crashes or coverage to assess these techniques, \TESTS4PY{} showcases the ability of test generators to identify functional bugs.
Despite the \emph{oracle problem} representing a challenge in automatically categorizing the generated tests, \TESTS4PY{} offers more profound insight into the effectiveness of test generators.
This approach could pave the way for a new breed of generators specifically targeting functional defects.
The benchmark also presents a variety of real-world faults, allowing for the evaluation of test generation techniques across different bug types and analyzing their effectiveness in uncovering and diagnosing these issues.
Given the capability of \TESTS4PY{} to generate tests, it provides ample material to study and learn from when setting up a test generator---for example, when employing symbolic execution.

\subsection{Mining Input Grammars}%
\label{sec-mining}

The included input grammar allows to validate input specification mining approaches and serves as a ground truth to calculate precision and recall for the derived specification, i.e., how many inputs derived from the specification are correct according to this ground truth grammar and how many correct inputs derived from this grammar are accepted by the mined specification.

\subsection{Driving Automatic Program Repair}%
\label{sec-apr}

The \TESTS4PY{} benchmark holds great potential for enhancing automatic program repair (APR) methodologies.
APR is an exciting area of research focused on devising techniques and tools that autonomously rectify software bugs.
Typically, APR strategies hinge on pinpointing the root cause of a defect, generating a fix, and validating it through test leveraging.

With its capacity to generate new tests, \TESTS4PY{} offers APR a dynamic platform to probe how different test sets featuring varying attributes impact APR.
Users can generate diverse test sets of different sizes and proportions of failing and passing tests and then apply APR to generate patches.
The accuracy of these generated solutions can then be evaluated using a concealed test set.
This investigation can provide insights into the properties a test set must possess to yield adequate repairs through APR.

Additionally, the subject-specific oracles provided in \TESTS4PY{} can be used to evaluate the synergistic combination of APR and test generation.
Existing research, such as the work by Yang et al.~\cite{yang2017opad}, has already integrated test generation and APR to minimize overfitting, albeit relying on weaker oracles.
However, our benchmark enables the combination of APR and functional test generation. Further exploration in this direction could significantly advance the research in this domain.

\subsection{Improving Automated Debugging}%
\label{sec-ad}

\TESTS4PY{} also holds significant potential for refining automated debugging techniques, most of which depend on the size and quality of a test set.
The benchmark's ability to \emph{generate new tests as needed} opens the door for deeper investigation into the requirements of a test set to unveil the fault-causing statements in a program.
As \TESTS4PY{} includes a patch for each subject and the possibility of retrieving the faulty statements from it, assessing the accuracy of the identified statements is a straightforward task.
Some debugging techniques already integrate test generation to refine their hypotheses, for example, \ALHAZEN{}~\cite{kampmann2020alhazen} or \AVICENNA{}~\cite{eberlein2023avicenna}.
These capabilities make \TESTS4PY{} a benchmark of choice for evaluating such methodologies in the context of real-world functional bugs.
Moreover, \TESTS4PY{} provides an embedded statistical fault localization with the integration of \SFLKIT{}~\cite{smytzek2022sflkit} that enables further research in this direction of automated debugging.

\section{Threats to Validity}%
\label{sec-ttv}

For each bug in our benchmark, we investigated its causes to ensure the quality of the created test cases and the test generation.
However, we may include tests that pass or fail for reasons other than the underlying defect.
To counter this threat, we verified that all tests (included or generated) pass or fail based on the oracle according to their labels.

Even though we tried to stay as close as possible to the underlying defects, we may implement code that does not reveal the original bug.
To minimize this risk, we verified the execution and oracles of all created test cases for correctness.

\section{Related Work}%
\label{sec-related}

\TESTS4PY{} bears significant parallels to \BUGSINPY{}~\cite{widyasari2020bugsinpy}, a source of inspiration and foundation for our work.
\BUGSINPY{} contributed significantly as the pioneer benchmark of real-world faulty Python programs.
Another Python program benchmark is \REFACTORY{}~\cite{hu2019refactoring}, based on student assignments and designed to evaluate automated program repair.
In the context of automated program repair, additional benchmarks such as \CODEFLAWS{}~\cite{tan2017codeflaws} for C and \BEARS{}~\cite{madeiral2019bears} for Java programs exist.
The \BUGSWARM{}~\cite{tomassi2019bugswarm} benchmark comprises faulty programs and their fixes for Python and Java programs.

While we focus on Python, benchmarks are available for several other programming languages.
A prominent example is \DEFECTS4J{}~\cite{rene2014defects4j}, a benchmark for faulty Java programs.
Another benchmark for buggy Java programs is \BUGSJAR{}~\cite{saha2018bugsjar}, which comprises a staggering~1,158 subjects.

In test generation, Google's \FUZZBENCH{}~\cite{metzman2021fuzzbench} stands out as a popular benchmark for evaluating and comparing fuzz testing based on the achieved coverage.
In debugging, we want to showcase the work by Böhme et al.~\cite{boehme2017dbgbench}.
Their efforts culminated in a benchmark where human experts analyzed defects and provided a diagnosis for each of the included subjects.

\section{Conclusion and Future Work}%
\label{sec-conclusion}

We introduce \TESTS4PY{}, a benchmark of real-world Python bugs. Each bug in this benchmark is accompanied by a test oracle and the capability to generate and execute system and unit tests.
\TESTS4PY establishes an easy-to-use, readily integrable architecture ready for everyday use.

Our future work will focus on the following topics:
First, we continue expanding our benchmark by including even more bugs from the \BUGSINPY{} database---eventually assimilating all subjects into \TESTS4PY{}.
We are also designing two studies to explore the impact of various test set properties on automatic program repair and statistical fault localization, as discussed in \Cref{sec-tg,sec-apr}.

\TESTS4PY{} is available as open source under
\begin{center}
    \href{https://github.com/smythi93/Tests4Py}{https://github.com/smythi93/Tests4Py}
\end{center}

\begin{acks}
  This research was partially funded by the Deutsche Forschungs- gemeinschaft (DFG, German Research Foundation) – GR 3634/4-2 Emperor (261444241). 
\end{acks}

\AtBeginEnvironment{thebibliography}{\balance}
\bibliographystyle{ACM-Reference-Format}
\bibliography{tests4py}

\end{document}